\documentclass[runningheads]{llncs}
\usepackage{graphicx}
\usepackage{pseudocode}
\usepackage{float}
\usepackage{amsfonts}
\usepackage[british]{babel}
\usepackage{color}
\usepackage{soul}
\usepackage{subfig}

\begin{document}
\title{Dyslexia detection from EEG signals using SSA component correlation and Convolutional Neural Networks}
\titlerunning{Dyslexia detection by SSA-CNN using EEG signals}
\author{Andr\'es Ortiz\inst{1} \and
Francisco J. Mart\'inez-Murcia\inst{1} \and
Marco A. Formoso\inst{1} \and
Juan Luis Luque\inst{2}\and
Auxiliadora S\'anchez\inst{2}
}
\authorrunning{A. Ortiz et al.}
\institute{Department of Communications Engineering, University of Malaga (Spain) \and Department of Developmental Psychology, University of Malaga (Spain)
\email{aortiz@ic.uma.es}}
\maketitle              % typeset the header of the contribution
\begin{abstract}
Objective dyslexia diagnosis is not a straighforward task since it is traditionally performed by means of the intepretation of different 
behavioural tests. Moreover, these tests are only applicable to readers. This way, early diagnosis requires the use of specific tasks not 
only related to reading. Thus, the use of Electroencephalography (EEG) constitutes an alternative for an objective and early diagnosis 
that can be used with pre-readers. In this way, the extraction of relevant features in EEG signals results crucial for classification. 
However, the identification of the most relevant features is not straighforward, and predefined statistics in the time or frequency domain 
are not always discriminant enough. On the other hand, classical processing of EEG signals based on extracting EEG bands frequency 
descriptors, usually make some assumptions on the raw signals that could cause indormation loosing. In this work we propose an alternative 
for analysis in the frequency domain based on Singluar Spectrum Analysis (SSA) to split the raw signal into components representing 
different oscillatory modes. Moreover, correlation matrices obtained for each component among EEG channels are classfied using a 
Convolutional Neural network.    

\end{abstract}
\section{Introduction}
Developmental dyslexia (DD) is a difficulty in the acquisition of reading skills, whose prevalence is estimated between 5\% and 12\% of the 
population \cite{Peterson2012}. It has an important social impact, since it may determine school failure and has harmful effects in the 
self-esteem of affected children. Prevention programs and individualized intervention tasks may help to mitigate behavioural aspects in 
dyslexic children, when they are applied in the early stages. Nevertheless, early diagnosis is currently a challenging task since most 
behavioural tests developed to this end include reading or writing tasks. 

Alternatively, the use of biomedical signals directed to measure brain activity, constitutes a powerful tool to develop differential and 
objective diagnosis methods. These signals can be acquired under a specific experimental setup that may not neccesary require any 
action from the subject.
In this work we use Electroencephalography (EEG) acquired during a non-interactive task consisting of the application of auditory 
stimuli that resemble the sampling processes performed in the brain for language processing. This segmentation proccess aims to extract 
features for recognising patterns related to different phonemes, sylabes or words. Thus, in this work we propose a method for the extraction 
of EEG features to be used in differential diagnosis. Moreover, these features may help to identify biomarkers to figure out unknown aspects 
of the DD related to its neural basis. This can offer valuable information for a better understanding of the differences between dyslexic 
and non-dyslexic subjects, with special application to the design of individualized intervention tasks \cite{Thompson2015}.

Usually, frequency features are used in EEG processing, specifically those related to the power distribution in different frequency bands 
(Delta, Theta, Alpha, Beta and Gamma). Frequency-based descriptors have been used in BCI and EVP experiments 
\cite{Ortega2016,Leon2019,Bradley2004}, by means of Fourier or Wavelet Analysis to estimate the average power in each band. These 
descriptors allowed to differentiate brain states or responses to diverse stimuli. As a matter of fact, different studies conducted 
in the search for DD-related patterns in EEG signals \cite{Power2016,Cutini2016} have shown differences in readers due 
to cognitive impairment of the phonological representation of word forms. Speech encoding which is related to speech prosody and 
sensorimotor synchronization problems can be revealed by finding patterns at different sub-bands.
In this work, we used EEG signals recorded by a 32 active electrodes BrainVision (Brain Products GmhH) equipment during 5 minute sessions, while presenting an auditive stimulus to the subject. These signals are then pre-processed and analyzed in the frequency domain by means of Singular Spectrum Analysis (SSA), which allows to decompose the raw signal into additive components representing different oscillatory modes.

The rest of the paper is organized as follows. Section \ref{sec:materialsmethods} presents details of the database and 
signal preprocessing. Then, section \ref{sec:methods} describes the auditory stimulus and the post-processing using SSA to extract 
features, as well as the classification method. Section \ref{sec:results} presents and discusses the 
classification results,  and finally, Section \ref{sec:conclusions} draws the main conclusions.

\section{Materials and methods}
\label{sec:materialsmethods}

\subsection{Database}
\label{sec:database}
The present experiment was carried out with the understanding and written consent of each child's legal guardian and in the presence 
thereof. Forty-eight participants took part in the present study, including 32 skilled readers (17 males) and 16 dyslexic readers (7 males) 
matched in age ($t(1) = -1.4, p > 0.05$, age range: 88-100 months). The mean age of the control group was $94,1\pm3.3$ months, and 
$95,6\pm2.9$ months for the dyslexic group. All participants were right‐handed Spanish native speakers with no hearing impairments and 
normal or corrected-‐to-‐normal vision. Dyslexic children in this study had all received a formal diagnosis of dyslexia in the school. None 
of the skilled readers reported reading or spelling difficulties or had received a previous formal diagnosis of dyslexia. Each subject was 
measured twice for each stimulus.

\section{Methods}
\label{sec:methods}
DD is a reading disorder characterized by reduced awareness of speech units \cite{Molinaro2016}. Recent models of neuronal speech 
coding suggest that dyslexia originates from the atypical dominant neuronal entrainment in the right hemisphere to the slow-rhythmic 
prosodic (Delta band, 0.5-4 Hz), syllabic (Theta band, 4-8 Hz) or the phoneme (Gamma band, 12-40 Hz), speech modulations, which are defined 
by the onset time (i.e., the envelope) generated by the speech rhythm \cite{Flanagan2018,Diliberto2018}.

According to \cite{Flanagan2018,Diliberto2018}, different brain rhythms involved in language processing are associated to neural 
oscillations that control the sampling processes developed to split auditory stimulus into its constituent parts, neccesary to recognise patterns at 
phoneme, syllabe and word. These neurophysiological responses should explain the manifestations of the temporal processing deficits 
described in dyslexia. In this work, EEG signals were obtained for auditory stimulus consisting in amplitude modulated (AM) white-noise at the rate of 8Hz. EEG signals were acquired at a sampling rate of 500 Hz, using a 32 active electrodes (BrainProducts actiCAP) while presenting the auditory stimulus.

\subsection{Signal preprocessing}
EEG signals were pre-processed in order to remove artefacts related to eye blinking and impedance variations due to movements. Since eye 
blinking signal is recorded along with EEG signals, these artefact are removed by blind source separation using Independent 
Component Analysis (ICA) \cite{Li2006}. Then, EEG signal of each channel is normalized independently to zero mean and unit 
variance to unify the scale in the subsequently processing carried out (for instance, Power Spectral Density calculation). In addition, a number of samples at the beginning and at the end of 
the signals were removed in order to obtain the same number of samples for each subject. Specifically, 136 seconds of EEG recording per 
subject and for each experiment were stored. It is worth noting that these 136 seconds signals were split into 40 seconds segments to speed up the post processing (such as SSA computation). Moreover, 40 seconds is enough time to capture the lowest EEG frequency considered (corresponding to Delta band ([0.5-4] Hz)) with a reasonable frequency resolution. Finally, all segments are band-pass filtered to keep only the frequencies of interest ([0.5, 40] Hz). Each segment is processed and used independently to generate samples for training the classifier.

\subsection{Singular Spectrum Analysis}
\label{sec:SSA}
Singular Spectrum Analysis (SSA) is a non-parametric spectral estimation method that decomposes the original time series into a sum of $K$ 
series. Formally, a time series $\{ X=x_{1}, x_{2},...,x_{N}\}$ is embedded into a vector space of dimension $\mathbb{K}$, composed of 
the eigenvectors of the covariance matrix $C_{x}$ computed for the L-lagged vectors. Lagged vectors $\overline{X_{i}}$ are defined as:

\begin{equation}
 \overline{X_{i}}=\{x_{i},...,x_{i+L-1}\} \in \mathbb{R}^L
\end{equation}

\noindent where $K=N-L+1$.
Thus, the covariance matrix :

\begin{equation}
 X_{x}=\frac{1}{N-L} \sum_{t=1}^{N-L} \overline{X(t)}\overline{X(t+L)}
\end{equation}

The $K$ eigenvectors $E_{k}$ of the lag-covariance matrix $C_{x}$ are called temporal empirical orthogonal functions (EOFs). Moreover, 
the eigenvalues $\lambda_{k}$ corresponding to each eigenvector account for the contribution of the direction $E_{k}$ to total variance.
This way, the projection of the original time series on the k-component can be computed as:

\begin{equation}
 Y_{k}(t)=\overline{X}(t+n-1)E_{k}(n)
\end{equation}

\begin{figure}[]
  \centering
  \includegraphics[width=0.8\textwidth]{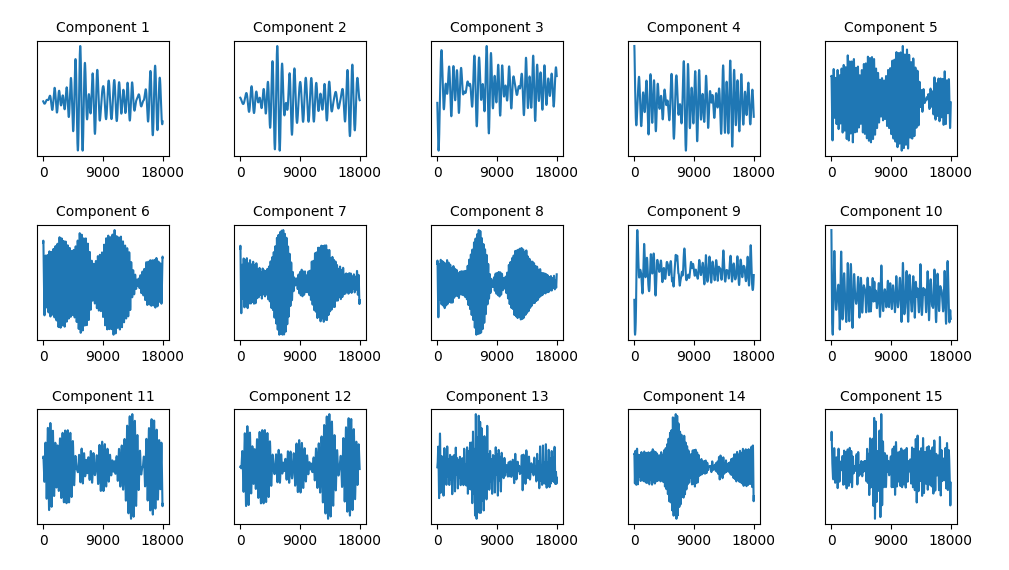}
  \caption{Example of 15 first SSA components extracted for channel 0}
  \label{fig:SSA_components}
\end{figure}

Each component extracted from SSA accounts for a part of the variance of the original signal. As these components are sorted according to 
the associated eigenvalues (in descending order), the first components will account for more variance than later components. Later, in 
Section \ref{sec:results}, we show the variance explained by the computed components.

On the other hand, as can be seen in Figure \ref{fig:SSA_components}, some of the extracted components are highly correlated. These correlated components can be grouped without losing interpretability. In this work, correlated EOFs were grouped to finally compose 5 components.  

\subsection{Feature extraction by Component Correlation}
\label{sec:channel_correlation}
The feature extraction stage in this work consists in the computation of the Pearson's correlation between channels for the PSD of each SSA component. This way, the PSD of each component is estimated using a modification of the Welch's method 
\cite{Welch1967}, a robust estimator that improves the standard periodogram by reducing the noise, but at the cost of reducing the spectral resolution. To perform the original method, the signal is divided into different segments overlapping semgents. Then, a modified periodogram is computed for each windowed segment, and the resulting periodograms are averaged. In our case, since a number of EEG segments are available per subject and electrode, we compute the modified periodogram over every segment. Here, the \textit{Hanning} window is used, and then the average periodogram is used to compute the correlation matrices.

\begin{figure}
  \centering
  \subfloat[]{\label{fig:cormats_CN}\includegraphics[width=0.9\textwidth]{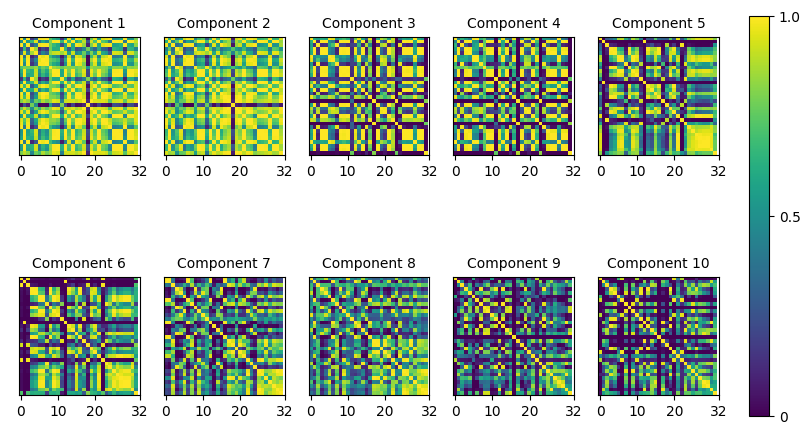}} 
  \\
  \subfloat[]{\label{fig:cormats_DD}\includegraphics[width=0.9\textwidth]{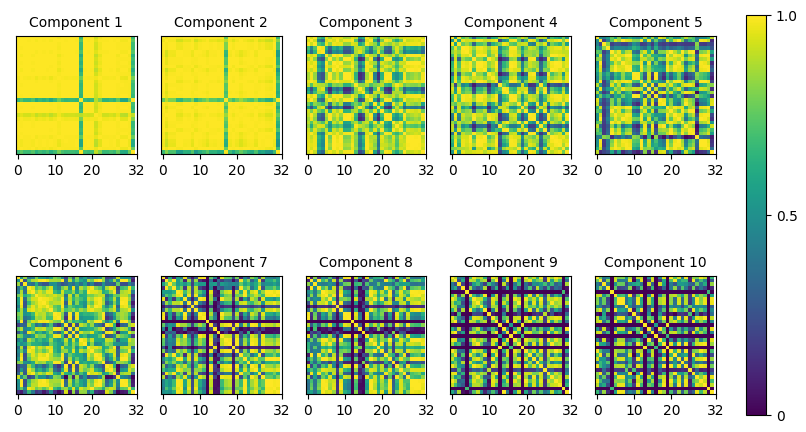}}
  \caption{Example of correlation matrices between channels for the first 10 SSA components computed for Control (a) and DD (b) subjects.}
  \label{fig:cormats}
\end{figure}

As shown in Figure \ref{fig:cormats}, there appear SSA components presenting very similar channel correlation matrices. This is the result of the computation of PSDs of highly correlated SSA components, as their PSD profile is very similar. These components can be grouped together since it indicates poorly separated components. On the contrary, well separated components generally exhibit low correlation. In fact, it is usual to group highly correlated components (in the time domain) by means of the so called \textit{weighted correlation matrix} (w-correlation), which allows identifying correlated components. 

This way, highly correlated SSA components are grouped together in the time domain and then, the PSD of the group is computed again. 
Since SSA produces additive components, they can be grouped by simply adding them. The resulting correlation matrices are then classified using a Convolutional Neural Network.

\subsection{Classification using an ensemble of CNNs}
\label{sec:CNN}
Once the correlation matrices for the PSD of the grouped components are computed, they are classified using a convolutional neural network. Convolutional neural networks are widely used to clasify image data with one or more channels (i.e. RGB images or even hyperspectral images) with important applications in  the Machine Learning community \cite{Krizhevsky2012,Ciresan2011,Ortiz2016}, especially within the artificial vision and image analysis fields.
CNNs are bioinspired by the convolutional response of neurons, and combine feature extraction and classification in one single 
architecture. The combination of different convolutional layers is able to recognize different patterns, from low-level features to higher 
abstractions, depending on the net depth. The set of fully connected layers (dense), similar to a perceptron, placed after convolutional 
layers, the classification. On the other hand, all neurons in any convolutional layer share the same weights, saving memory and easing the computation of the convolutions. This considerably reduces the number of trainable parameters in the network in comparison to a perceptron-like network designed for the same classification task.

CNN architectures have evolved over time, including new layers and connections among layers that have outperformed previous 
image classification approaches. An example of improvement for CNN is the use of residual blocks \cite{He2016}, consisting in adding to the input of a convolutional layer the input of a previous convolutional layer. Other improvements include the use of batch normalization layers or the use of strides$>$1 in convolutional layers instead of poolings.   

\begin{figure}[]
  \centering
  \includegraphics[width=0.7\textwidth]{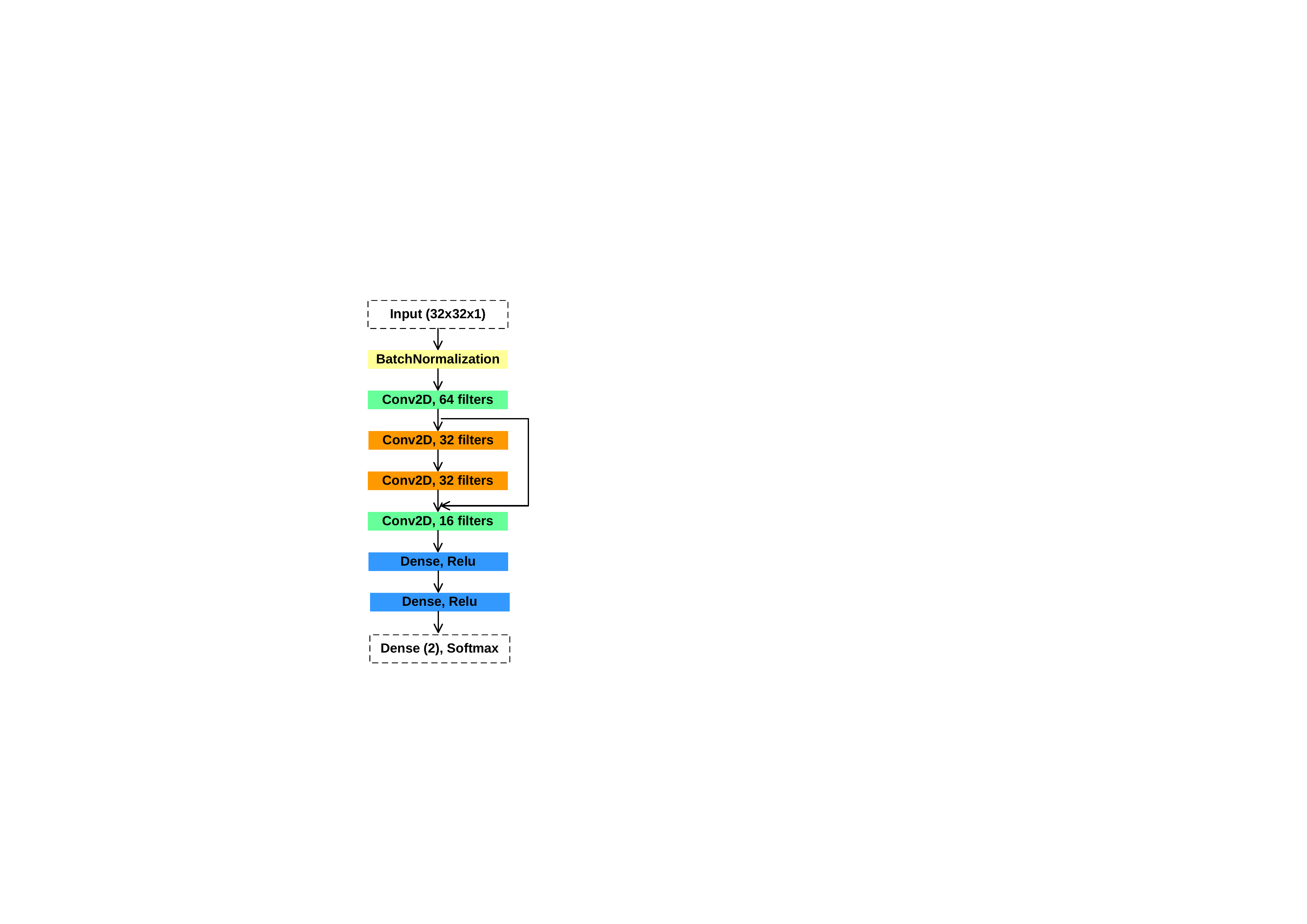}
  \caption{Architecture of the Convolutional Neural Network used for classifying subjects by means of their channel correlation 
matrices}
  \label{fig:CNN}
\end{figure}

In this work, a CNNs are used to classify the subjects by means of their channel correlation matrices explained in Section 
\ref{sec:channel_correlation}. Since a matrix correlation is computed for each grouped component, we used a CNN with the architecture shown in Figure \ref{fig:CNN} for each one. Subsequently, the output of all CNN-based classifiers are combined by a majority voting 
strategy. The use of ensebles of CNNs have demonstrated their effectivity in image classification tasks, increasing the performance 
obtained by a single CNN while diminishing the overfitting effects \cite{Krizhevsky2012}.

\section{Experimental Results}
\label{sec:results}
\begin{figure}
  \centering
  \subfloat[]{\label{fig:varexp}\includegraphics[width=0.45\textwidth]{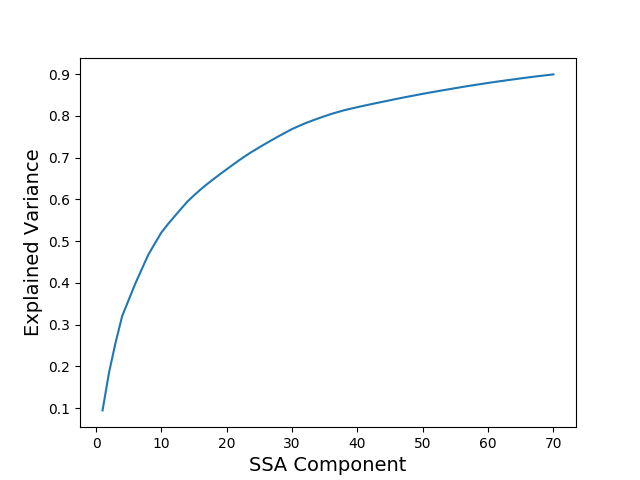}} 
  \hspace{0.1cm}
  \subfloat[]{\label{fig:grouped_varexp}\includegraphics[width=0.45\textwidth]{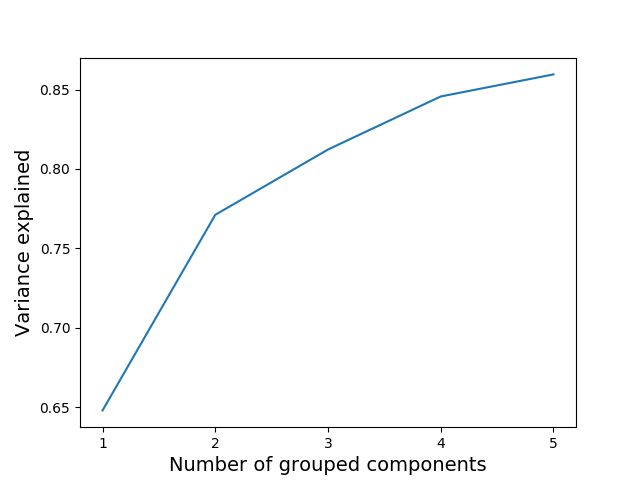}}
  \caption{Variance explained by SSA (a) individual components, (b) grouped components (5 groups)}
  \label{fig:variance_explained}
\end{figure}

In this section, we show the results of the classification experiments carried out to demonstrate the discriminative capabilities of the  features extracted from EEG signals. As explained in Section \ref{sec:methods}, the input to each of the CNNs consists of a channel 
correlation matrix for a specific grouped component. Moreover, since components showing a high correlation between them are guessed to belong to the same source, they can be added up. This way, 5 clusters in the \textit{w-correlation} matrix have been computed by the hierarchical clustering of the pairwise distances among entries in the w-correlation matrix. Indeed, Figure \ref{fig:varexp} show the variance explained by 70 components and Figure \ref{fig:grouped_varexp} the variance explained by the 5 groups of components.

Grouping components has two main advantages. On the one hand, the classifier architecture is simpler. On the other hand, components with lower variance that could cause missclassifications (as they are not informative enough) are added together in a group explaining a larger part of the variance than those individual components.

\begin{figure}[]
  \centering
  \includegraphics[width=0.7\textwidth]{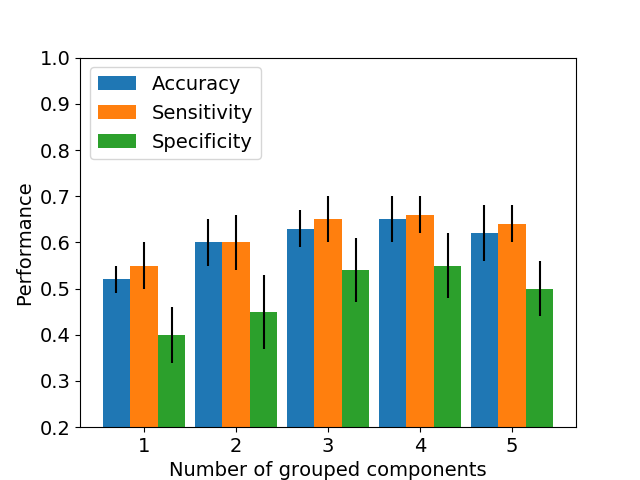}
  \caption{Classification performance obtanied for different number of grouped components. Accuracy, Sensitivity and Specificity are shown}
  \label{fig:clasif_performance}
\end{figure}

Figure \ref{fig:clasif_performance} show the classification performance obtained when a different number of grouped components are considered. As shown, the use of 4 grouped components provide the best performance. It is worth noting that 3 components (for instance) in Figure \ref{fig:clasif_performance} indicates that an ensemble composed of 3 CNNs is used.

As shown in Figure \ref{fig:clasif_performance} (a), since grouped components are sorted indescending order of the variance explained, first grouped components explain the most part of the variance, and therefore, provide the best classffication results. However, the last grouped component are most likely to contain noisy components, causing missclassifications.

\section{Conclusions and future work}
\label{sec:conclusions}
In this work we present a classification method for EEG signals based on SSA and CNN. The core idea in this work is the use of between 
channel correlation of the PSD of different oscillatory modes contained in the EEG signals. These oscillatory modes are represented by SSA components. An analysis of the components was performed and those showing higher correlations are grouped to compose a reduced set of components that accumulates a higher part of the variance.  The use of grouped components reduces the number of CNN stages in the ensemble (since one CNN per component is used) and removes highly correlated components from the input of the classifier, improving the performance.
The approach shown in this work shows its effectivity for extracting informative features for the differential diagnosis of DD. 
As a future work, we plan to continue exploring the SSA components searching for single oscillatory modes in EEG channels, represented by 
the fundamental frequency of a component or from harmonics, that could be synchronized among channels. This way, instead of using the 
correlation matrix, a matrix containing a synchronizaton index among channels could be used. This also have important and interpretable 
biological implications regarding the identificaton of cooperative brain areas - and the analysis of differences in the cooperation between 
controls and DD subjects - in language processing task.   

\section*{Acknowledments}
This work was partly supported by the MINECO/FEDER under PGC2018-098813-B-C32 project. We gratefully acknowledge the support of NVIDIA Corporation with the donation of one of the GPUs used for this research. Work by F.J.M.M. was supported by the MICINN ``Juan de la Cierva - Formaci\'on'' Fellowship. We also thank the \textit{Leeduca} research group and Junta de Andalucía for the data supplied and the support.


\end{document}